%%%%%%%%%%%%%%%%%%%%%%%%%%%%%%%%%%%%%%%%%%%%%%%%%%%%%%%%%%%%%%%%%%%%%%%%%%%%%%%%%%
% Setup
%%%%%%%%%%%%%%%%%%%%%%%%%%%%%%%%%%%%%%%%%%%%%%%%%%%%%%%%%%%%%%%%%%%%%%%%%%%%%%%%%%

%%	JNL.TEX					Doug Eardley
%%
\message
{JNL.TEX version 0.92 as of 6/9/87.  Report bugs and problems to Doug Eardley.}
%%
%%	This is a set of TeX 82 macros designed to produce scientific
%%	papers with a minimum of fuss and using as much of plain.tex as
%%	possible.  The user need only know what is in the TeXbook, and
%%	the macros under ``user definitions'' below.  Also, the user
%%	definitions are intended to be as simple as possible, so that
%%	the user may change them as desired.  I have tried to avoid all
%%	cleverness, although it may have snuck in here and there.
%%
%%	A considerable degree of compatibility with AmSTeX is maintained,
%%	although not guaranteed.  The intention is that AmSTeX input file
%%	should run with only a few changes near the beginning;  see
%%	discussion below under "AmSTeX compatability".
%%
%%	For documentation, see the file JNLHLP.TEX.  Optional features are
%%	contained in the files PPT.TEX (for two-up preprints), REFORDER.TEX
%%	(automatic numbering of references), EQNORDER.TEX (automatic numbering
%%	of equations), and TABLEOFC.TEC (automatic generation of table of
%%	contents).

%%
%%	Redefine \input to prevent files being loaded more than once
%%
\catcode`@=11
\expandafter\ifx\csname inp@t\endcsname\relax\let\inp@t=\input
\def\input#1 {\expandafter\ifx\csname #1IsLoaded\endcsname\relax
\inp@t#1%
\expandafter\def\csname #1IsLoaded\endcsname{(#1 was previously loaded)}
\else\message{\csname #1IsLoaded\endcsname}\fi}\fi
\catcode`@=12

%%
%%  Font definitions suitable for the IMAGEN-480 (Written by Tony Kennedy)
%%

%  Define a whole menagerie of pseudo-12pt fonts

%\font\twelverm=amr10 scaled 1200    \font\twelvei=ammi10 scaled 1200
%\font\twelvesy=amsy10 scaled 1200   \font\twelveex=amex10 scaled 1200
%\font\twelvebf=ambx10 scaled 1200   \font\twelvesl=amsl10 scaled 1200
%\font\twelvett=amtt10 scaled 1200   \font\twelveit=amti10 scaled 1200
%\font\twelvesc=amcsc10 scaled 1200  \font\twelvesf=amssmc10 scaled 1200
%\skewchar\twelvei='177   \skewchar\twelvesy='60

%  Define \...point macros to change fonts and spacings consistently

%	tenpoint

%%
%%	Various internal macros
%%

\def\beginlinemode{\endmode
  \begingroup\parskip=0pt \obeylines\def\\{\par}\def\endmode{\par\endgroup}}
\def\beginparmode{\endmode
  \begingroup \def\endmode{\par\endgroup}}
\let\endmode=\par
{\obeylines\gdef\
{}}
\def\singlespace{\baselineskip=\normalbaselineskip}

\def\oneandahalfspace{\baselineskip=\normalbaselineskip
  \multiply\baselineskip by 3 \divide\baselineskip by 2}
\def\doublespace{\baselineskip=\normalbaselineskip \multiply\baselineskip by 2}

%%
%%	Page layout, margins, font and spacing (feel free to change)
%%

%\hsize=6.5truein
%\hoffset=1truein
%\vsize=8.9truein
%\voffset=1truein
%\parskip=\medskipamount
%\def\\{\cr}
%\twelvepoint		% selects twelvepoint fonts (cf. \tenpoint)
%\doublespace		% selects double spacing for main part of paper (cf.
			%	\singlespace, \oneandahalfspace)
%\overfullrule=0pt	% delete the nasty little black boxes for overfull box

%%
%%	The user definitions for major parts of a paper (feel free to change)
%%

  %  "Draft", Timestamp

	% Preprint number at upper right of title page

\def\title			%  Title on title page
  {\null\vskip 3pt plus 0.2fill
   \beginlinemode \doublespace \raggedcenter \bf}

\def\author			%  Author(s) name(s)  on title page
  {\vskip 3pt plus 0.2fill \beginlinemode
   \singlespace \raggedcenter\sc}

\def\affil			% Affiliations (can intermix with \author)
  {\vskip 3pt plus 0.1fill \beginlinemode
   \oneandahalfspace \raggedcenter \sl}

\def\abstract			% Begin abstract
  {\vskip 3pt plus 0.3fill \beginparmode
   \oneandahalfspace ABSTRACT: }

\def\endtitlepage		% End title page, begin body of paper
  {\endpage			% 	This subsumes \body
   \body}

\def\body			% Begin text body;  can be used to end
  {\beginparmode}		% \title, \author, \affil, \abstract,
				% \reference, or \figurecaption modes

\def\beginitems{
\par\medskip\bgroup\def\i##1 {\item{##1}}\def\ii##1 {\itemitem{##1}}
\leftskip=36pt\parskip=0pt}
\def\enditems{\par\egroup}

\def\beneathrel#1\under#2{\mathrel{\mathop{#2}\limits_{#1}}}

\def\refto#1{$^{#1}$}		% For references in text as superscript

\def\references			% Begin references -- basic format is Phys Rev
  {%\head{References}		% I.e., volume, page, year (space after commas).
   \beginparmode
   \frenchspacing \parindent=0pt \leftskip=1truecm
   \parskip=8pt plus 3pt \everypar{\hangindent=\parindent}}

\gdef\refis#1{\item{#1.\ }}			% Ref list numbers.

\gdef\journal#1, #2, #3, 1#4#5#6{		% Journal reference.  Comma sets
    {\sl #1~}{\bf #2}, #3 (1#4#5#6)}		% off: name, vol, page, year

\def\endreferences{\body}

\catcode`@=11
\newcount\r@fcount \r@fcount=0
\newcount\r@fcurr
\immediate\newwrite\reffile
\newif\ifr@ffile\r@ffilefalse
\def\w@rnwrite#1{\ifr@ffile\immediate\write\reffile{#1}\fi\message{#1}}

\def\writer@f#1>>{}
\def\referencefile{%			  Stuff to write .REF file
  \r@ffiletrue\immediate\openout\reffile=\jobname.ref%
  \def\writer@f##1>>{\ifr@ffile\immediate\write\reffile%
    {\noexpand\refis{##1} = \csname r@fnum##1\endcsname = %
     \expandafter\expandafter\expandafter\strip@t\expandafter%
     \meaning\csname r@ftext\csname r@fnum##1\endcsname\endcsname}\fi}%
  \def\strip@t##1>>{}}

\def\citeall#1{\xdef#1##1{#1{\noexpand\cite{##1}}}}
\def\cite#1{\each@rg\citer@nge{#1}}	% Variable No. of args, separated by ","

\def\each@rg#1#2{{\let\thecsname=#1\expandafter\first@rg#2,\end,}}
\def\first@rg#1,{\thecsname{#1}\apply@rg}	% each@ag is a general purpose
\def\apply@rg#1,{\ifx\end#1\let\next=\relax%	  variable no. of arg. macro.
\else,\thecsname{#1}\let\next=\apply@rg\fi\next}% args separated by commas

\def\citer@nge#1{\citedor@nge#1-\end-}	% Check for M-N range (M and N numbers)
\def\citer@ngeat#1\end-{#1}
\def\citedor@nge#1-#2-{\ifx\end#2\r@featspace#1 % Single argument
  \else\citel@@p{#1}{#2}\citer@ngeat\fi}	% M-N range of arguments
\def\citel@@p#1#2{\ifnum#1>#2{\errmessage{Reference range #1-#2\space is bad.}%
    \errhelp{If you cite a series of references by the notation M-N, then M and
    N must be integers, and N must be greater than or equal to M.}}\else%
 {\count0=#1\count1=#2\advance\count1 by1\relax\expandafter\r@fcite\the\count0,%
  \loop\advance\count0 by1\relax%	  Loop from M to N
    \ifnum\count0<\count1,\expandafter\r@fcite\the\count0,%
  \repeat}\fi}

\def\r@featspace#1#2 {\r@fcite#1#2,}	% Eat spaces at beginning or end of arg
\def\r@fcite#1,{\ifuncit@d{#1}%		  Cite individual reference
    \newr@f{#1}%
    \expandafter\gdef\csname r@ftext\number\r@fcount\endcsname%
                     {\message{Reference #1 to be supplied.}%
                      \writer@f#1>>#1 to be supplied.\par}%
 \fi%
 \csname r@fnum#1\endcsname}
\def\ifuncit@d#1{\expandafter\ifx\csname r@fnum#1\endcsname\relax}%
\def\newr@f#1{\global\advance\r@fcount by1%
    \expandafter\xdef\csname r@fnum#1\endcsname{\number\r@fcount}}

\let\r@fis=\refis			% Save old \refis, redefine
\def\refis#1#2#3\par{\ifuncit@d{#1}%      Use two params #2 #3 to strip blank
   \newr@f{#1}%
   \w@rnwrite{Reference #1=\number\r@fcount\space is not cited up to now.}\fi%
  \expandafter\gdef\csname r@ftext\csname r@fnum#1\endcsname\endcsname%
  {\writer@f#1>>#2#3\par}}

\def\ignoreuncited{%   redefine \refis if ignoring uncited references
   \def\refis##1##2##3\par{\ifuncit@d{##1}%
     \else\expandafter\gdef\csname r@ftext\csname r@fnum##1\endcsname\endcsname%
     {\writer@f##1>>##2##3\par}\fi}}

\def\r@ferr{\endreferences\errmessage{I was expecting to see
\noexpand\endreferences before now;  I have inserted it here.}}
\let\r@ferences=\references
\def\references{\r@ferences\def\endmode{\r@ferr\par\endgroup}}

\let\endr@ferences=\endreferences
\def\endreferences{\r@fcurr=0%		  Save old \endreferences, redefine
  {\loop\ifnum\r@fcurr<\r@fcount%	  Loop over refnum and produce text
    \advance\r@fcurr by 1\relax\expandafter\r@fis\expandafter{\number\r@fcurr}%
    \csname r@ftext\number\r@fcurr\endcsname%
  \repeat}\gdef\r@ferr{}\endr@ferences}

% Save old \endpaper, redefine it to write parting message.

\let\r@fend=\endpaper\gdef\endpaper{\ifr@ffile
\immediate\write16{Cross References written on []\jobname.REF.}\fi\r@fend}

\catcode`@=12

\citeall\refto		% These macros will generate citations
%\citeall\ref		%
%\citeall\Ref		%

\input epsf.tex
\magnification 1200
\vsize=7.8in
\hsize=5.7in
\voffset=0.1in
\hoffset=-0.15in
\newcount\eqnumber
\baselineskip 18pt plus 0pt minus 0pt

%%%%%%%%%%%%%%%%%%%%%%%%%%%%%%%%%%%%%%%%%%%%%%%%%%%%%%%%%%%%%%%%%%%%%%%%%%%%%%%%%%
% Fonts
%%%%%%%%%%%%%%%%%%%%%%%%%%%%%%%%%%%%%%%%%%%%%%%%%%%%%%%%%%%%%%%%%%%%%%%%%%%%%%%%%%

\font\rmmthree=cmbx10 scaled 1500
\font\rmmtwo=cmbx10 scaled 1200
\font\rmmoneB=cmbx10 scaled 1100

\font\rmmoneI=cmti10 scaled 1000

\font\ninerm=cmr10 scaled 900

\font\eightrm=cmr10 scaled 800
\font\eightbf=cmbx10 scaled 800

%%%%%%%%%%%%%%%%%%%%%%%%%%%%%%%%%%%%%%%%%%%%%%%%%%%%%%%%%%%%%%%%%%%%%%%%%%%%%%%%%%
% Formatting Macros
%%%%%%%%%%%%%%%%%%%%%%%%%%%%%%%%%%%%%%%%%%%%%%%%%%%%%%%%%%%%%%%%%%%%%%%%%%%%%%%%%%

\def\title#1{\centerline{\noindent{\rmmthree #1}}\nobreak\smallskip\eqnumber=1}

\def\nosectbegin#1{\bigskip\bigbreak\leftline{\rmmtwo #1}\nobreak\medskip}

%%%%%%%%%%%%%%%%%%%%%%%%%%%%%%%%%%%%%%%%%%%%%%%%%%%%%%%%%%%%%%%%%%%%%%%%%%%%%%%%%%
% Various Macros
%%%%%%%%%%%%%%%%%%%%%%%%%%%%%%%%%%%%%%%%%%%%%%%%%%%%%%%%%%%%%%%%%%%%%%%%%%%%%%%%%%

\def\lapp{\hbox{$ {     \lower.40ex\hbox{$<$}
                   \atop \raise.20ex\hbox{$\sim$}
                   }     $}  }
\def\gapp{\hbox{$ {     \lower.40ex\hbox{$>$}
                   \atop \raise.20ex\hbox{$\sim$}
                   }     $}  }

\def\marbul{\strut\vadjust{\kern-2pt$\bullet$}}

\def\specialwarn{\vtop to
\strutdepth{\baselineskip\strutdepth\vss\llap{
\lower.1ex\hbox{$\bigtriangleup$}\kern-0.884em$\triangle$\kern-0.5667em{\eightrm
!}\hskip 13.5pt}\null}}
\def\strutdepth{\dp\strutbox}

%%%%%%%%%%%%%%%%%%%%%%%%%%%%%%%%%%%%%%%%%%%%%%%%%%%%%%%%%%%%%%%%%%%%%%%%%%%%%%%%%%
% Equation Numbering
%%%%%%%%%%%%%%%%%%%%%%%%%%%%%%%%%%%%%%%%%%%%%%%%%%%%%%%%%%%%%%%%%%%%%%%%%%%%%%%%%%

\def\new{{\the\eqnumber}\global\advance\eqnumber by 1}
\def\delaynew{{\the\eqnumber}}
\def\nownew{\global\advance\eqnumber by 1}
\def\last{\advance\eqnumber by -1 {\the\eqnumber}
    \global\advance\eqnumber by 1}
\def\eqnam#1{%%%Naming macro
\xdef#1{\the\eqnumber}}

%%%%%%%%%%%%%%%%%%%%%%%%%%%%%%%%%%%%%%%%%%%%%%%%%%%%%%%%%%%%%%%%%%%%%%%%%%%%%%%%%%
% Figure Macros
%%%%%%%%%%%%%%%%%%%%%%%%%%%%%%%%%%%%%%%%%%%%%%%%%%%%%%%%%%%%%%%%%%%%%%%%%%%%%%%%%%

\def\figure#1#2#3#4#5#6{
\topinsert
\null\hskip #4\relax
\epsfxsize #2
\epsfysize #3
\epsfbox{#1}
\medskip
{\baselineskip 10pt\noindent\narrower\rm\hbox{\eightbf
Figure #5}:\quad\eightrm
#6 \smallskip}
\endinsert
}

%%%%%%%%%%%%%%%%%%%%%%%%%%%%%%%%%%%%%%%%%%%%%%%%%%%%%%%%%%%%%%%%%%%%%%%%%%%%%%%%%%
% Specfic Macros
%%%%%%%%%%%%%%%%%%%%%%%%%%%%%%%%%%%%%%%%%%%%%%%%%%%%%%%%%%%%%%%%%%%%%%%%%%%%%%%%%%

\def\dalemb#1#2{{\vbox{\hrule height .#2pt
\hbox{\vrule width.#2pt height#1pt \kern#1pt\vrule width.#2pt}
\hrule height.#2pt}}}

\def\tdot{\kern -8.5pt {}^{{}^{\hbox{...}}}}
\def\dotprime{\kern -8.0pt{}^{{}^{\hbox{.}~\prime}}}

\def\foot#1#2{\footnote{#1}{\eightrm #2}}

%%%%%%%%%%%%%%%%%%%%%%%%%%%%%%%%%%%%%%%%%%%%%%%%%%%%%%%%%%%%%%%%%%%%%%%%%%%%%%%%%%
% Document
%%%%%%%%%%%%%%%%%%%%%%%%%%%%%%%%%%%%%%%%%%%%%%%%%%%%%%%%%%%%%%%%%%%%%%%%%%%%%%%%%%

\voffset 1in
\baselineskip 12pt
\vskip 100pt
\title{PRIMORDIAL GRAVITATIONAL WAVES:}
\medskip
\title{A PROBE OF THE VERY EARLY UNIVERSE}
\vskip 30pt
\centerline{\rmmoneB R.$\,$A. Battye}
\vskip 10pt
\centerline{\rmmoneI Theoretical Physics Group, Blackett Laboratory}
\centerline{\rmmoneI Imperial College of Science, Technology \& Medicine}
\centerline{\rmmoneI University of London}
\centerline{\rmmoneI Prince Consort Road, London SW7 2BZ, U.K.}
\vskip 10pt
\centerline{and}
\vskip 10pt
\centerline{\rmmoneB E.$\,$P.$\,$S. Shellard}
\vskip 10pt
\centerline{\rmmoneI Relativity and Gravitation Group}
\centerline{\rmmoneI Department of Applied Mathematics and Theoretical Physics}
\centerline{\rmmoneI University of Cambridge}
\centerline{\rmmoneI Silver Street, Cambridge CB3 9EW, U.K}
\vskip 40pt
\baselineskip 12pt
\medskip \centerline{\rmmoneB Summary} 
\medskip 
{\narrower{\baselineskip 9pt \ninerm \noindent
We discuss the potential cosmological role of gravitational wave astronomy 
as a probe of the very early universe. 
The next generation of detectors---now in production---may be able to observe
a stochastic background of gravitational waves produced by violent
processes during the earliest moments after the creation of the universe.  
Viable theoretical scenarios within detector sensititivity
include strongly first-order phase transitions, possibly at the end 
of inflation, and networks of cosmic strings. 
At this stage, other primordial backgrounds from slow-roll 
inflation, global topological defects and the standard electroweak phase 
transition appear to be out of range. The discovery of any of these possible
cosmological sources will have enormous implications for our understanding 
of the very early universe and for fundamental physics at the highest energies.
}\smallskip}   
\vskip 80pt
\centerline{\rmmoneI Submitted to the Gravity Research Foundation essay 
competition}
\centerline{\rmmoneI  29th March, 1996}
\vfill\eject
\voffset 0.1in
\pageno=1
\baselineskip 18pt

\noindent The discovery of the cosmic microwave background radiation was a 
watershed in modern cosmology.  The more recent observations of 
fluctuations in this background provide us with a detailed snapshot of 
the universe at about 400,000 years after the Hot Big Bang, just as
the universe became transparent to electromagnetic
radiation.  The scientific impact of the discovery of the CMBR is difficult to
overestimate, yet it serves merely as a useful foil in the context of this
essay.   Here, we wish to discuss the prospect of gravitational wave
astronomy providing snapshots, not from a few hundred thousand years,
but from the very first fractions of a second after the creation of the
universe.  Gravitational radiation easily penetrates
the electromagnetic surface of last scattering and propagates freely to
the present day from its time of emission, which can be as early as the
Planck epoch at $10^{-43}$ seconds (as illustrated in fig.~1).  

The remarkable transparency of the universe to gravitons is due to their 
very weak coupling with ordinary
matter which makes them extremely difficult to observe. In recent years, pioneering experiments have been proposed to directly 
detect gravitational waves using terrestrial and space-based laser 
interferometers\refto{thorne}. These experiments have been primarily 
designed to search for  bursts of radiation from astrophysical 
sources, such as black hole--black hole mergers, but they also have the 
potential to detect a stochastic background of gravitational waves produced 
by violent processes in the very early universe. Such background noise 
might at first be an annoyance for experimentalists, as it was for
Penzias and Wilson, but its implications would be far-reaching because the
processes responsible for its creation would undoubtedly involve physics beyond 
the standard model.  Currently, a number of viable theoretical models
produce detectable gravitational wave backgrounds, including first-order
phase transitions, topological defects, and inflation (refer to fig.~2),
but we must not preclude the strong possibility of the emergence of
some fundamentally new physical
insight; after all, serendipity and observational cosmology seem to be habitual
associates. 
Nevertheless, we are not suggesting that this exciting, but somewhat 
speculative, cosmological scenario should override
the solid astrophysical case which has already been made 
for the next generation of interferometers.
Rather our aim is to emphasise, with some specific examples, the potential 
rewards of such a unique probe of the very early universe. 

\figure{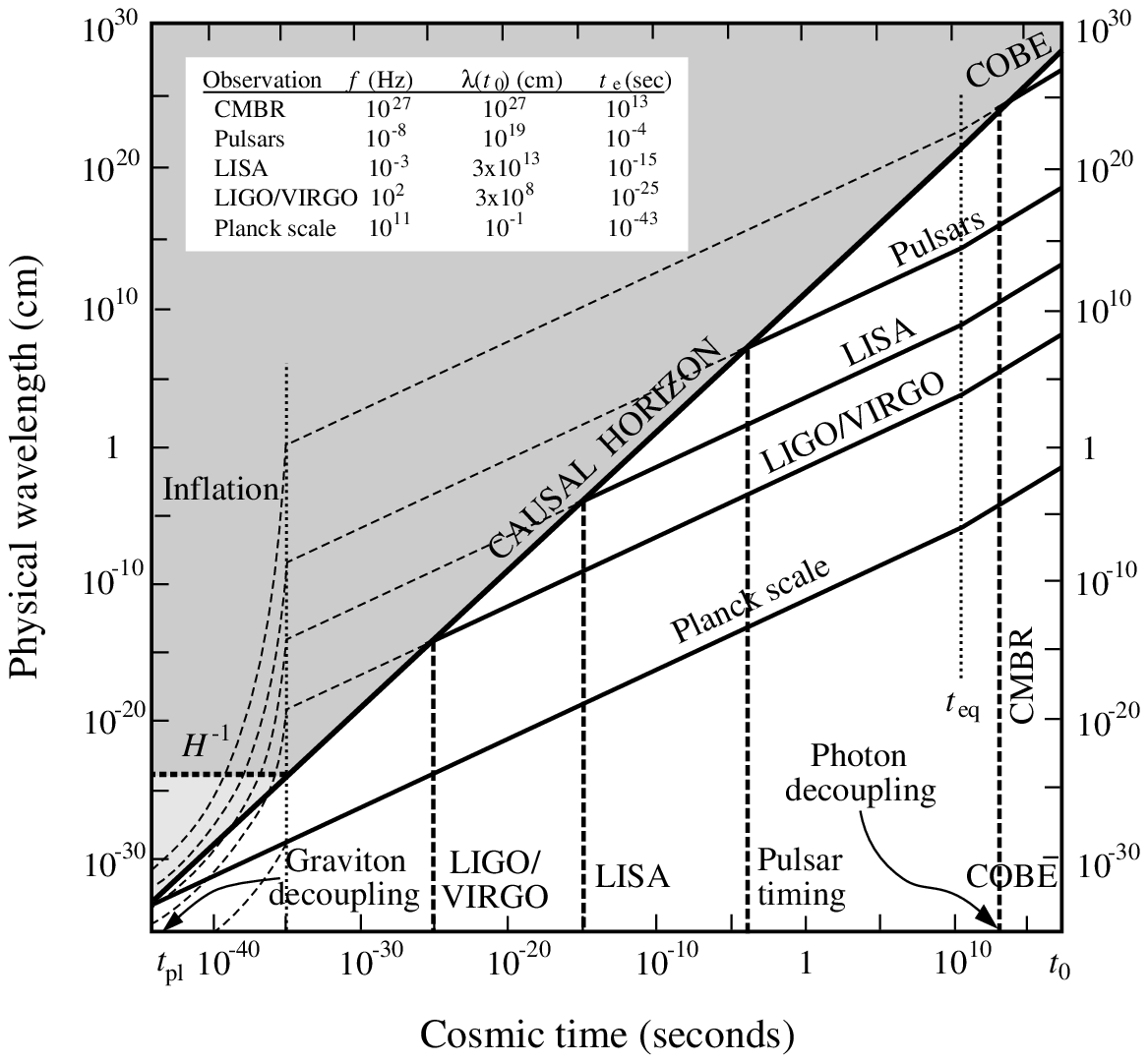}{4.5in}{4.0in}{0.0in}{1}{A summary of the times probed 
by the various gravitational wave experiments. For classical, causal sources 
emitting sub-horizon 
scale radiation, the time of emission is calculated by redshifting the 
wavelength today back to the causal horizon at emission. For gravitational
radiation produced during inflation (the dashed lines), this procedure gives 
the time at which 
a particular wavelength comes back inside the horizon.}

Stochastic backgrounds of gravitational waves are normally quantified by
their relative spectral density $\Omega_{\rm g}(f)$ given at a
frequency $f$, that is, the energy density in gravitational radiation in 
an octave
frequency bin centred on $f$ relative to the critical density of the universe.
This is directly related to the dimensionless wave amplitude 
($h_c \propto \sqrt{\Omega_g}/f$) which is
measured experimentally, except that we must allow for cosmological
uncertainties by quoting sensitivities in terms of $\Omega_g h^2$, where
`little $h$' is the rescaled Hubble parameter ($0.4<h<0.9$).  At present, 
there are four main frequency bands for studying 
gravitational radiation (each shown in fig.~2).  First, there is the 
tensor contribution to the
microwave background anisotropies detected by the COBE-DMR experiment\refto{COBE};
this implies an upper bound of $\Omega_{\rm g}<7\times 10^{-11}$ at 
frequencies around $3\times 10^{-17}h\,{\rm Hz}$.  A second upper limit comes
from pulsar observations\refto{kaspi}, since a stochastic background would lead 
to timing noise in the incoming periodic signal; this imposes the tighter 
constraint, 
$\Omega_{\rm g}h^2<6\times 10^{-8}$ at $4\times 10^{-9}{\rm Hz}$\foot{*}{It should be noted that the 
statistical veracity of this result has been
 questioned by recent re-analyses of the same 
data suggesting both stronger\refto{thorsett} and weaker\refto{mchugh}. 
bounds. For the purposes
of this essay we shall use the original work\refto{kaspi}, though simple 
scaling can be used 
to accommodate any readjustment.}.  Thirdly, proposed
ground-based interferometers will study frequencies around 100Hz with a
maximum sensitivity of about $\Omega_{\rm g}h^2\approx 10^{-7}$ for the 
first LIGO detector\refto{LIGO} and $\Omega_{\rm g}h^2\approx 10^{-10}$ 
for the advanced LIGO and VIRGO\refto{VIRGO} detectors.
Finally, the ambitious space-based interferometer LISA\refto{LISA} will have  
a sensitivity of $\Omega_{\rm g}h^2\approx 10^{-10}$ at about $10^{-3}{\rm Hz}$.
 
\figure{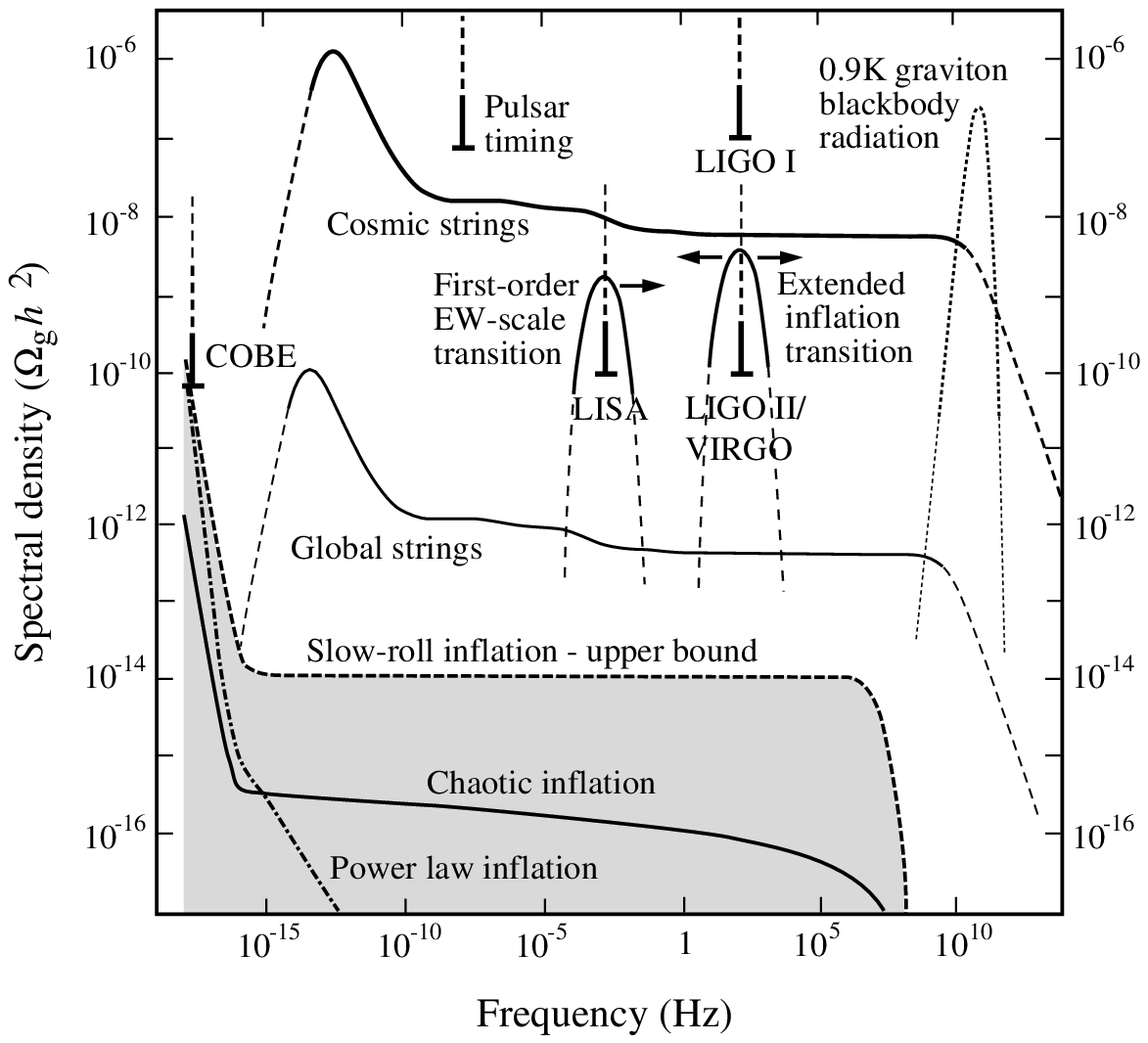}{4.5in}{4.0in}{0.0in}{2}{Summary of the potential cosmological sources of a stochastic
gravitational radiation background, including inflationary models, 
first-order phase transitions and cosmic strings\refto{AC,BCSa}, as well as a primordial 0.9K 
blackbody graviton spectrum (the analogue of the blackbody photon radiation). 
Also plotted are the relevant constraints from the COBE measurements, 
pulsar timings, and the sensitivities of the proposed interferometers. 
Notice that local cosmic strings and strongly first-order phase transitions 
may produce detectable backgrounds, in contrast to standard slow-roll 
inflation models.}

Gravitational radiation produced by classical, causal processes in the 
early universe will have a characteristic frequency related to its
time of emission since the causal horizon provides an upper limit on the
wavelength. Figure 1 illustrates the implications of this for
the LIGO and VIRGO detectors which 
remarkably probe times as early as 
$10^{-25}$ seconds, while LISA potentially looks back to $10^{-15}$ seconds.

First, however, we focus attention on the quantum creation of 
gravitational waves
during an inflationary epoch.  Inflation is a period of rapid expansion which 
occurs if the potential energy of a scalar field
dominates the energy density of the early universe. Inflationary scenarios 
are popular because they
resolve the well-known horizon, flatness and monopole problems of the standard 
cosmology. 
However, probably their most significant testable prediction 
is the quantum mechanical production of a nearly scale-invariant spectrum of 
adiabatic density perturbations.   An important by-product of this process 
is the parametric amplification of tensor modes resulting in a background of 
gravitational waves.

\figure{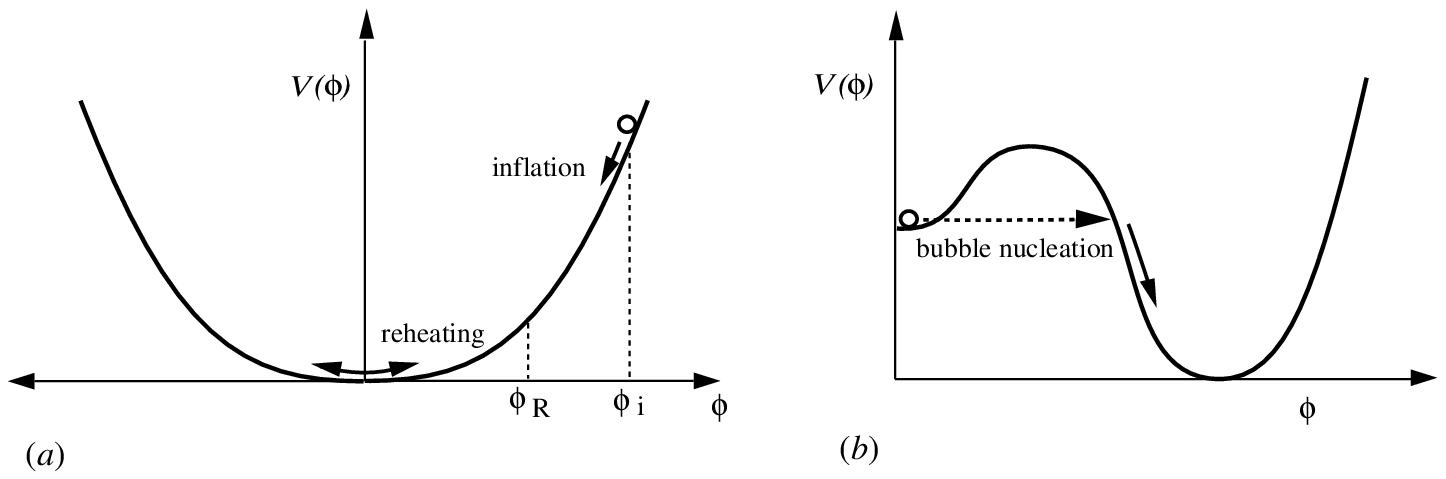}{5.0in}{1.55in}{0.0in}{3}{(a) Scalar field potential 
$V(\phi)$ for chaotic inflation.  Inflation occurs while the field $\phi$ 
slowly rolls down the
potential and ends in reheating when $\phi$ finally 
oscillates about the bottom of the potential. 
(b) A potential giving rise to a first-order phase transition.  
The field $\phi$ becomes trapped in the metastable false vacuum state 
and then escapes 
through quantum tunnelling or large thermal fluctuations.}

The standard picture for inflation is that the scalar field rolls slowly down 
its potential as shown in fig.~3(a), finally ending in the re-heating of the 
universe when
the remaining vacuum energy is converted into radiation. During this slow-roll 
regime the amplitude of the spectrum of gravitational waves produced is 
proportional to $H^2$, where $H^{-1}$ is the Hubble radius. After creation at 
about $10^{-35}$ seconds, 
a particular mode is driven outside the Hubble radius (or `horizon') by the 
rapid expansion and it effectively `freezes' until it returns inside the 
horizon during the subsequent radiation or matter dominated eras. This process
is illustrated by the dashed `super-horizon' evolution in fig.~1.

The important dynamical fact is that $H$ must decrease monotonically and 
therefore the largest contribution to the gravitational wave spectrum is 
due to modes that were driven outside the horizon early in inflation and 
have just come back inside the horizon at the present day. Fig.~1 shows 
that these are the  frequencies probed by CMBR experiments and the 
observed anisotropies can be used to normalize $H$ early in 
inflation. Once this fact is known, the entire spectrum can be calculated
using the dynamical equations for the scalar field. Assuming that the 
entire COBE signal is due to tensor modes, a weak upper 
bound on the inflationary signal can be deduced, that is,
$\Omega_{\rm g}h^{2}< 10^{-14}$ for frequencies which 
came inside the horizon before equal matter-radiation.  This  
is well below the quoted sensitivities of proposed detectors.

In fact, for any realistic slow-roll inflation model the situation is 
considerably worse and, paradoxically, the larger the CMBR tensor contribution, the 
lower the contribution to the frequencies relevant for direct detection; $H$ 
must decrease more rapidly in models which have a large gravitational wave 
contribution in COBE frequencies\refto{liddle}.
The actual spectrum for two simple inflationary models is illustrated in 
fig.~2.  A simple chaotic model has a negligible COBE signal but an 
almost flat spectrum with amplitude $\Omega_{\rm g}h^2\approx 10^{-16}$ 
at a few Hz.  In contrast, for a power law inflation model, almost all the 
COBE signal is due to gravitational waves 
but the spectrum falls very rapidly to an amplitude of only 
$\Omega_{\rm g}h^2\approx 10^{-24}$ by the LISA frequency 
band.

While the prospects for detecting gravitational waves from standard
slow-roll inflation seem poor, there 
are a number of alternative scenarios which can produce a detectable 
signal. These include speculative superstring-inspired 
models\refto{superstring} and also those which do not end in the standard 
re-heating scenario. Extended inflation\refto{extend} and 
hybrid inflation\refto{hybrid} exit through a phase transition which 
provides an extra, potentially more powerful source of gravitational 
waves if this final transition is strongly first-order\refto{TW}. 

A simple potential for a first-order phase transition is illustrated in 
fig.~3(b); the field becomes trapped in a metastable local minimum of the 
potential---the false vacuum---before the transition to the true vacuum 
takes place by bubble nucleation and growth (as shown in fig.~4).
When these bubbles collide copious amounts of 
gravitational radiation can be emitted, particularly if the relative 
velocity of the bubble walls is highly relativistic as in a strongly first-order
phase transition. For an extended inflation model, the contribution 
to the gravitational wave 
background from such a transition could be as high as 
$\Omega_{\rm g}h^2\approx 10^{-8}-10^{-9}$ and it would lie in the 
LIGO/VIRGO sensitivity range if the re-heat 
temperature of the universe was close to $10^{8}$GeV or $10^{9}$GeV.  
We have included a crude sample spectrum for such a model in fig.~2.

\figure{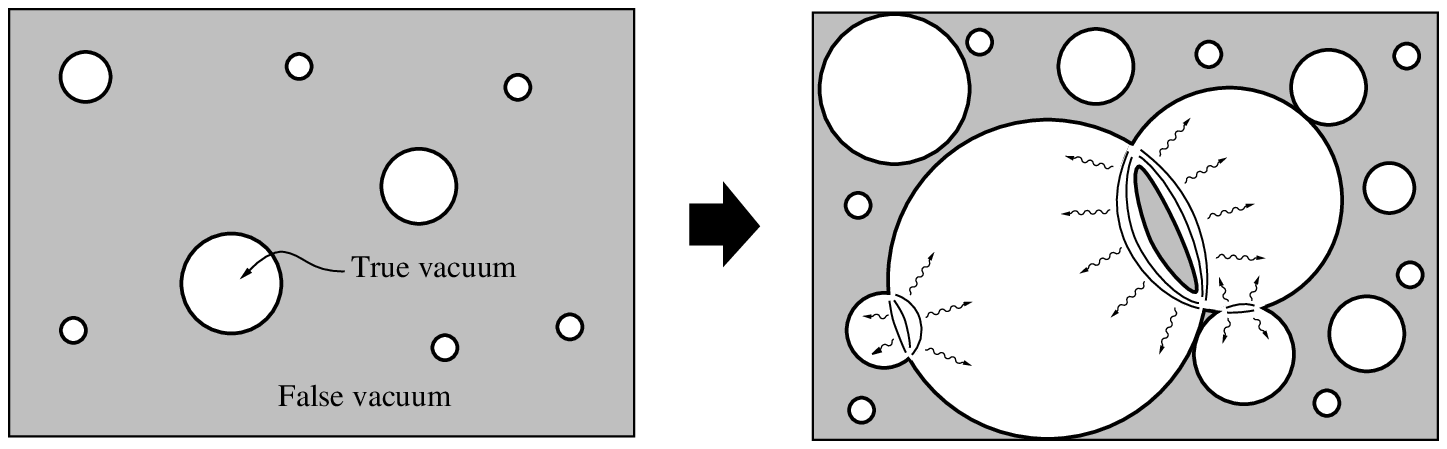}{5.0in}{1.63in}{0.0in}{4}{A strongly first-order
phase transition can produce a stochastic background of gravitational
waves.  The nucleated bubble walls expand at highly relativistic velocities and
complete the transition to the true vacuum through violent collisions.
}

First- and second-order phase transitions are a generic phenomena in cosmology 
since spontaneously 
broken symmetries are now an integral part of modern particle physics.
If the standard electroweak phase transition were strongly first-order, 
then there would be a contribution of about $\Omega_{\rm g}h^2\approx 10^{-9}$ 
around a frequency of $10^{-3}$Hz, inside the LISA band with a detectable 
amplitude\refto{KTW} (see fig.~2). Unfortunately, the minimal standard model 
is currently believed to have only a weakly first-order transition with the
bubble walls reaching velocities well below the speed of light; estimates  
in this case,\refto{KKT} yield only $\Omega_{\rm g}h^2\sim 10^{-22}$ .  
Nevertheless, there are a number
of well-motivated extensions to the standard model (like supersymmetry) 
which could entail symmetry-breaking just above the electroweak scale; 
if strongly first-order, such phase transitions would create a 
distinctive and detectable LISA signal.  Fortuitously from this point of view, 
LISA has a very interesting frequency response range.

Symmetry breaking phase transitions in the early universe will also inevitably 
produce topological defects of one form or another (refer to the 
recent review\refto{VS}).  Of particular 
interest in our context are line-like defects, 
known as cosmic strings, formed at a grand-unification scale ($10^{-35}$ seconds). 
Due to their large  mass per unit length (typically 
$\mu\sim 10^{22}{\rm g\,cm}^{-1}$), such strings provide a viable mechanism
for seeding the formation of large-scale 
structure\refto{structureone,structuretwo}. A string network has been 
shown in numerical simulations\refto{BB,AS}
to evolve towards a self-similar scaling regime in which the number of strings
per horizon volume remains fixed (see fig.~5).  The network maintains this 
constant relative density by creating loops which oscillate relativistically 
and decay radiatively.  For local cosmic
strings, this loop decay channel is gravitational waves, thus creating a 
stochastic background over a vast range of frequencies from $10^{-13}$Hz to 
$10^{10}$Hz.

Under mild assumptions about the loop emission spectrum and the
particle content of the universe,
the radiation background due to loops created in the radiation era has a
flat spectrum with amplitude $\Omega_{\rm g}h^2\approx 10^{-8}$ 
for frequencies between $10^{-9}$Hz and $10^{10}$Hz.
In contrast, the background produced in the matter era is sensitive to the 
loop spectrum. If this spectrum is not 
cut-off by radiation backreaction\refto{BSc}, then loop radiation can
 feed into higher $(\sim 10^{-8}{\rm Hz})$ frequencies, because 
it does not feel the full impact of the redshifting after 
the time of equal matter-radiation\refto{AC}.  Interestingly, 
it is precisely these frequencies that are relevant for the pulsar timing 
experiments\refto{kaspi,thorsett,mchugh}. Using an appropriately truncated 
loop spectrum and the dimensionless string parameter normalised to COBE\refto{ACSSV},
$G\mu/c^2\approx 10^{-6}$, the predicted radiation background can be 
calculated numerically\refto{BCSa} as shown in fig.~2. 
Alternatively, if one assumes that $G\mu/c^2$ is arbitrary, then the pulsar 
timing experiment\refto{kaspi} can be used to deduce the bound
$G\mu/c^2<3.5(\pm 0.8)\times 10^{-6}$.

\figure{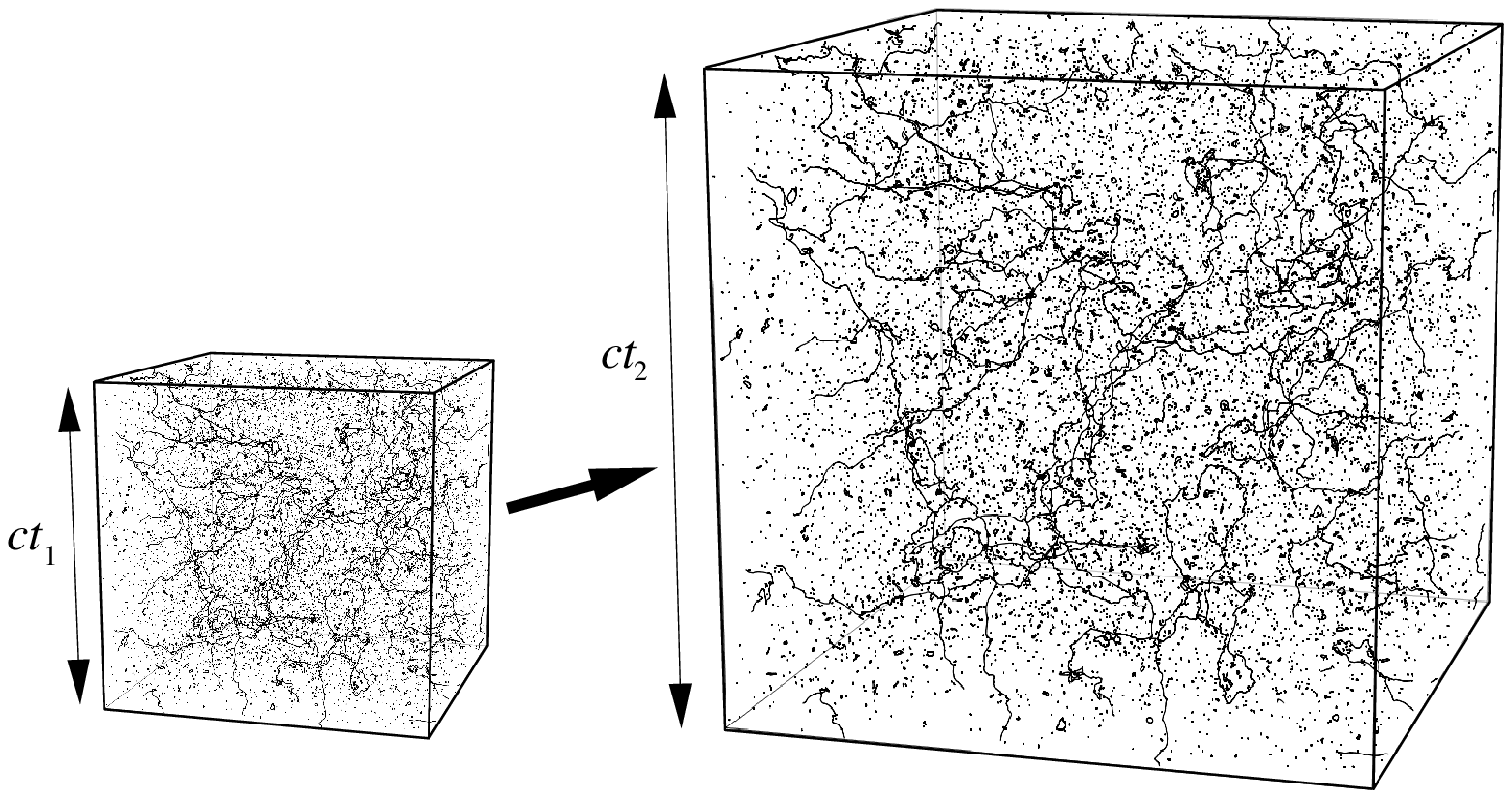}{4.17in}{2.6in}{0.0in}{5}{The scale-invariant evolution 
of a string network is maintained by small loop creation; the oscillating 
loops then decay into a stochastic background of gravitational waves. One 
horizon volume of a radiation era simulation is shown containing about 
40 long strings\refto{AS}.}

The contribution from cosmic strings to the frequency ranges relevant for 
direct detection
is likely to be slightly lower than that for pulsar timing. This is due to 
particle mass thresholds, where the number of relativistic degrees of freedom 
${\cal N}$ decrease, effectively diluting the relative 
contribution of any pre-existing decoupled radiation. 
The cosmic string spectrum shown in 
fig.~2 illustrates this effect with a gradual rise 
at frequencies around $10^{-4}$Hz. This is caused by 
particle annihilation near the QCD and electroweak phase transitions. If the 
particle physics model has more degrees of freedom at higher energies there 
could be other steps associated with other phase transitions. 
However, the dependence on ${\cal N}$ is reasonably weak and therefore we 
can conservatively estimate that  $\Omega_{\rm g}h^2>5.0\times 10^{-9}$ 
at $10^{-3}$Hz and $\Omega 
_{\rm g}h^2>1.0\times 10^{-9}$ at 100Hz for $G\mu/c^2\approx 1.0\times 10^{-6}$. 
We note also that very precise determinations of the 
stochastic background from cosmic strings at different frequencies would 
measure $\cal N$ in different cosmological epochs, providing 
fascinating insight into the particle content of the early universe (at times
much earlier than the electroweak phase transition using LIGO or VIRGO).

Finally, we should note that there are other types of topological defects
which do not 
produce such a large gravitational wave background, notably those formed 
when global symmetries are broken.  For global cosmic strings, which primarily 
radiate Goldstone bosons, a simple calculation suggests that 
gravitational radiation will be suppressed by approximately 
four orders of magnitude relative to local strings, as illustrated in fig.~2.
It is likely that a background produced by other global defects, 
such as global textures, will be likewise suppressed.

We have summarized the potential cosmological sources of gravitational waves 
and concluded that, of the candidates already proposed, strongly first-order
phase transitions, possibly at the end of inflation, and local 
cosmic string networks provide tangible hope of direct detection.  On the other hand,
slow-roll inflation, global topological defects and the 
standard electroweak phase transition create signals which appear to be 
too weak. However, it is important to note that the study of gravitational wave 
emission in the early universe is a relatively unexplored domain.  In
light of the dramatic technological advances being made by
experimentalists, the time is clearly ripe for more detailed
quantitative studies of the characteristic signals produced by
cosmological sources.  Moreover, such theoretical foresight could
positively influence observational strategies.

To conclude, then, it is clear that gravitational waves 
can potentially provide a unique and unparallelled 
probe of the very early universe. The proposed interferometers, both terrestrial and
space-based, and improved pulsar timings could rule out or severely constrain 
a number of viable theoretical models. On the other hand,
the detection of a primordial background of gravitational waves would
have a profound impact on our understanding of high energy physics and 
cosmology, providing an unprecedented view of the earliest moments of our 
Universe.

%%%%%%%%%%%%%%%%%%%%%%%%%%%%%%%%%%%%%%%%%%%%%%%%%%%%%%%%%%%%%%%%%%%%%%%%%%%%%%%%%%
% References
%%%%%%%%%%%%%%%%%%%%%%%%%%%%%%%%%%%%%%%%%%%%%%%%%%%%%%%%%%%%%%%%%%%%%%%%%%%%%%%%%%

%%% PARAGRAPH SHAPE:

%\def\hang{\hangindent20pt\hangafter1\noindent}
\def\hang{}

%%% STANDARD JOURNAL:
%%% The following information must be supplied.
%%% \jnl{Authors|year|Article title|Journal|Volume|Page} 

\def\jnl#1#2#3#4#5#6{\hang{#1 [#2], {\it #4\/} {\bf #5}, #6.}
									}

%%% JOURNAL WITH ERRATUM:
%%% \jnlerr{Authors|year|Article title|Journal|Volume|Page|Err Volume|Err page} 

\def\jnlerr#1#2#3#4#5#6#7#8{\hang{#1 [#2], {\it #4\/} {\bf #5}, #6.
{Erratum:} {\it #4\/} {\bf #7}, #8.}
									}

%%% TWO JOURNALS:
%%% \jnltwo{Authors|year|Article title|Journal|Volume|Page|Journal|Vol.|Page} 

%%% PREPRINT:
%%% \prep{Authors|year|Article title|Number and comments} 
\def\prep#1#2#3#4{\hang{#1 [#2], #4.}
									}

%%% PROC:
%%% \proc{Authors|year|Article title|Book title|Editor|Publisher, City} 
\def\proc#1#2#3#4#5#6{\hang{#1 [#2], in {\it #4\/}, #5, eds.\ (#6).}
}

%%% BOOK:
%%% \book{Authors|year|Book title|Publisher, City (where applicable)} 
\def\book#1#2#3#4{\hang{#1 [#2], {\it #3\/} (#4).}
									}

%%% GENERAL REF:
%%% \genref{Author|year|Whatever} 

%%% Commonly used journals

\def\prl{Phys.\ Rev.\ Lett.}
\def\pr{Phys.\ Rev.}
\def\pl{Phys.\ Lett.}

\def\apj{Ap.\ J.}

\def\mn{M.$\,$N.$\,$R.$\,$A.$\,$S.}

\def\cupress{Cambridge University Press}

\def\wss{World Scientific, Singapore}

\nosectbegin{Bibliography}

\references

\baselineskip 10pt
\let\it=\nineit
\let\rm=\ninerm
\let\bf=\ninebf
\rm 

\def\cram{\vskip -5pt}

\refis{thorne}\proc{Thorne K.}{1996}{}{Particle and nuclear astrophysics and cosmology in the next millennium}{Kolb E.W \& Peccei R.}{\wss}\cram

\refis{kaspi}\jnl{Kaspi V.M., Taylor J.H. \& Ryba M.F.}{1994}{}{\apj}{428}{713}\cram

\refis{thorsett}\prep{Thorsett S.E. \& Dewey R.J.}{1996}{}{to appear 
{\it Phys.\ Rev. D}. }\cram

\refis{mchugh}\prep{McHugh M.P., Zalamansky G., Vernotte F. \& Lantz
E.}{1996}{}{submitted to {\it Phys.\ Rev. D}.}\cram

\refis{liddle}\jnlerr{Liddle A.}{1994}{}{\pr}{D49}{3805}{D51}{4603}\cram

\refis{superstring}\jnl{Brustein R., Gasperini M., Giovannini M. \& Veneziano G.}{1995}{}{\pl}{361B}{45}\cram

\refis{KTW}\jnl{Kowosky A., Turner M.S. \& Watkins R.}{1992}{}{\prl}{69}{2026}\cram

\refis{KKT}\jnl{Kamionkowski M., Kowosky A. \& Turner M.S.}{1994}{}{\pr}{D49}{2837}\cram

\refis{extend}\jnl{La D. \& Steinhardt P.J.}{1989}{}{\prl}{62}{376}\cram

\refis{hybrid}\jnl{Linde A.D.}{1994}{}{\pr}{49}{748}\cram

\refis{TW}\jnl{Turner M.S. \&  Wilczek F.}{1990}{}{\prl}{65}{3080}\cram

\refis{VS}\book{Vilenkin A. \& Shellard E.P.S.}{1994}{Cosmic strings 
and other topological defects}{\cupress}\cram

\refis{structureone}
\jnl{Zel'dovich Ya.B.}{1980}{}{\mn}{192}{663}\cram

\refis{structuretwo}
\jnlerr{Vilenkin A.}{1981}{}{\prl}{46}{1169}{46}{1496}\cram

\refis{BB}\jnl{Bennett D.P. \& Bouchet F.R.}{1990}{}{\pr}{D41}{2408}\cram

\refis{AS}\jnl{Allen B. \& Shellard E.P.S.}{1990}{}{\prl}{64}{119}\cram

\refis{BCSa}\prep{Battye R.A., Caldwell R.R. \& Shellard E.P.S}{1996}{}
{DAMTP preprint}\cram

\refis{BSc}\jnl{Battye R.A. \& Shellard E.P.S}{1996}{}{\pr}{D53}{1811}\cram

\refis{AC}\jnl{Caldwell R.R. \& Allen B.}{1992}{}{\pr}{D45}{3447}\cram

\refis{ACSSV}\prep{Allen B., Caldwell R.R., Shellard E.P.S., Stebbins A.
\& Veeraraghavan S.}{1995}{}{work in progress}\cram

\refis{COBE}\jnl{Davis R.L., Hodges H.M., Smoot G.F., Steinhardt P.J. \& Turner M.S.}{1992}{}{\prl}{69}{1856}\cram

\refis{LIGO}\jnl{Abramovici A. {\it et al}}{1992}{}{Science}{256}{325}\cram

\refis{VIRGO}\jnl{Bradachia {\it el al}}{1990}{}{Nucl. Instrum. \& Methods}{A289}{518}\cram

\refis{LISA}\jnl{Jafry Y.R., Cornelisse J. \& Reinhard R.}{1994}{}{ESA Journal}{18}{219}\cram

\endreferences

\vfill
\end